\documentclass[fleqn,12pt]{article}
\usepackage{graphics,epsfig,here,mathrsfs}

\def\mueg{\mu^- \to e^- \gamma}
\def\taueg{\tau^- \to e^- \gamma}
\def\taumug{\tau^- \to \mu^- \gamma}
\newcommand {\eq} [1] {Eq.~(\ref{#1})}
\newcommand {\fig} [1] {Fig.~\ref{#1}}

\def\lsim{\raise0.3ex\hbox{$\;<$\kern-0.75em\raise-1.1ex\hbox{$\sim\;$}}}
\def\gsim{\raise0.3ex\hbox{$\;>$\kern-0.75em\raise-1.1ex\hbox{$\sim\;$}}}

\newcommand{\AddrVienna}{
\it  Institut f\"ur Theoretische Physik, Universit\"at Wien, \\ 
A-1090 Vienna, Austria \\}

\newcommand{\AddrGAKUGEI}{%
 \it Department of Physics, Tokyo Gakugei University, Koganei, \\ 
Tokyo 184-8501, Japan\\}

\newcommand{\AddrHEPHY}{%
 \it Institut f\"ur Hochenergiephysik der \"Osterreichischen Akademie 
der Wissenschaften, A-1050, Austria\\}

\newcommand{\AddrAHEP}{%
 \it
Instituto de F\'\i sica Corpuscular / C.S.I.C.- Universitat
de Valencia, Edificio Institutos de Paterna, Apartado de Correos 22085,
E-46071 Valencia, Spain\\}

\newcommand{\AddrETH}{%
 \it Institut f\"ur Theoretische Physik, Universit\"at Z\"urich, \\ 
CH-8057 Z\"urich, Switzerland}

\begin{document}

\begin{flushright}
   TGU-35
 \\UWThPh-2005-13
 \\HEPHY-PUB 811/05
 \\IFIC/05-31
 \\ZU-TH 10/05
\end{flushright}

\begin{center}  
  \textbf{\large 
Test of lepton flavour violation at LHC}\\[10mm]

{A. Bartl${}^1$, K.~Hidaka${}^2$, K.~Hohenwarter-Sodek${}^1$, 
T. Kernreiter${}^1$, W.~Majerotto${}^3$ and W. Porod${}^{4,5}$
} 
\vspace{0.3cm}\\ 

$^1$ \AddrVienna
$^2$ \AddrGAKUGEI
$^3$  \AddrHEPHY
$^4$ \AddrAHEP
$^5$  \AddrETH
\end{center}

\noindent

\begin{abstract}
\noindent
We study lepton flavour violating decays of neutralinos
and sleptons within the Minimal Supersymmetric Standard Model, assuming
two and three generation mixings in the slepton sector.
We take into account the most recent bounds on flavour violating rare 
lepton decays. Taking the SPS1a' scenario as an example, we show that some of the 
lepton flavour violating branching ratios of neutralinos and sleptons can 
be sizable ($\sim$ 5-10\%).
We study the impact of the lepton flavour violating neutralino and slepton decays 
on the di-lepton mass distributions measured at LHC.
We find that they can result in novel and characteristic edge structures
in the distributions. 
In particular, double-edge structures can appear in the $e\tau$
and $\mu\tau$ mass spectra if $\tilde\tau_1$ is the lightest slepton.
The appearance of these remarkable structures provides a 
powerful test of supersymmetric lepton flavour
violation at LHC.
\end{abstract}

\newpage 

\section{Introduction}

There are stringent experimental constraints on lepton flavour violation
(LFV) in the charged lepton sector, the strongest coming from the
decay branching ratio of $\mueg$, BR$(\mueg) < 1.2 \times 10^{-11}$
\cite{Brooks:1999pu}. Others are BR$(\mu^- \to e^- e^+ e^-) <
10^{-12}$ \cite{Bellgardt:1987du}, BR$(\taueg) < 1.1 \times 10^{-7}$ 
\cite{Aubert:2005wa}, BR$(\taumug) < 6.8 \times 10^{-8}$ \cite{Aubert:2005ye},
BR$(\tau^-\to\mu^-\mu^+\mu^-) < 1.9 \times 10^{-7}$ \cite{Yusa:2004gm} 
and the limit on $\mu^- N\to e^- N$,
$R_{\mu e}< 7.8\times 10^{-13}$ \cite{muecon}, with
$R_{\mu e}=\Gamma[\mu^- N(Z,A)\to e^- N(Z,A)]/
\Gamma[\mu^- N(Z,A)\to \nu_\mu N(Z-1,A)]$.
In particular, the bounds on BR$(\taueg)$ and BR$(\taumug)$
have recently been substantially improved.

On the other hand, the various neutrino experiments
have clearly established that individual lepton flavour is violated 
(for a recent review see e.g. \cite{Maltoni:2004ei}).
In supersymmetric (SUSY) extensions of the Standard Model LFV
can originate in the slepton sector due to soft SUSY breaking
parameters, e.g. mass matrices with flavour off-diagonal entries.
Several studies along this line have been performed assuming 
either specific high-scale models
or specifying the LFV parameters at the low scale (see for instance 
\cite{photonex,modelle,collider,ref8}).

For the LHC it has been shown that SUSY LFV can be observed by studying
the LFV decays of the second neutralino 
$\tilde\chi^0_2$ arising from cascade decays of gluinos
and squarks, i.e. $\tilde\chi^0_2\to \tilde\ell \ell'\to \ell' \ell''\tilde\chi^0_1$:
signals of SUSY LFV can be extracted despite considerable
backgrounds and stringent experimental bounds on flavour violating lepton
decays in case of two generation mixings in either the right or left slepton
sector in the mSUGRA model 
\cite{Agashe:1999bm,ref12,Hisano:2002tk}. The $\tilde e_R-\tilde \mu_R$
mixing case was studied in \cite{Agashe:1999bm,Hisano:2002tk} and the
$\tilde \mu_L-\tilde \tau_L$ mixing case in \cite{ref12}.

In this paper we study the cases of two and three generation mixings 
in the slepton sector.
We take the SPS1a' point as a reference scenario and work out in detail
the individual branching ratios of the LFV two-body decays of the neutralino
and sleptons. An interesting feature of this scenario
is that the $\tilde\tau_1$ is lighter than the $\tilde e_R$ and 
$\tilde\mu_R$. As we will show, in the LFV case this leads to a novel 
double-edge structure in the $e\tau$ and $\mu\tau$ invariant
mass distributions, not appearing in the cases previously studied.

We take into account the contraints on the LFV parameters from
the most recent experimental limits on the rare decays $\ell^- \to \ell'^- \gamma$. 
This practically implies that the constraints from the rare three-body decays 
are fulfilled as they are dominated by virtual
photon exchange \cite{photonex}. The only possible exception is that from 
$\tau^- \to \mu^- \mu^+ \mu^-$ enhanced by 
the Higgs boson exchange for large $\tan\beta$ \cite{Dedes:2002rh}. 
This, however, does not apply to our case as we study a scenario with
$\tan\beta=10$ in our numerical analysis.
In the considered parameter range also the rate for
$\mu-e$ conversion is well below the corresponding experimental limit \cite{photonex}.

\section{The model}

The most general charged slepton mass matrix including left-right mixing
as well as flavour mixing in the basis of
$(\tilde e_L,\tilde\mu_L,\tilde\tau_L,\tilde e_R,\tilde\mu_R,\tilde\tau_R)\equiv 
(\tilde\ell_{1L},\tilde\ell_{2L},\tilde\ell_{3L},\tilde\ell_{1R},
\tilde\ell_{2R},\tilde\ell_{3R})$ 
is given by:
\begin{equation}
M^2_{\tilde \ell} = \left(
\begin{array}{cc}
M^2_{LL} &  M^{2\dagger}_{RL} \\
M^2_{RL} &  M^2_{RR} \\
\end{array} \right)~,
\label{eq:sleptonmass}
\end{equation}
where the entries are $3 \times 3$ matrices. They are given by
\begin{eqnarray}
\label{eq:massLL}
M^2_{LL,\alpha\beta} &=& 
M^2_{L,\alpha\beta} + \frac{ v^2_d Y^{E*}_{\alpha\gamma} Y^{E}_{\beta\gamma} }{2}
+ \frac{\left( {g'}^2 -  g^2 \right) (v^2_d - v^2_u) 
\delta_{\alpha\beta}}{8}  \, ,\\
\label{eq:sleptonmassLR}
M^2_{RL,\alpha\beta} &=& \frac{ v_d A_{\beta\alpha} 
- \mu^* v_u Y^E_{\beta\alpha} }{\sqrt{2}}  \, ,\\ 
M^2_{RR,\alpha\beta} &=& M^2_{E,\alpha\beta} 
+ \frac{ v^2_d Y^{E}_{\gamma\alpha} Y^{E*}_{\gamma\beta} }{2}
-\frac{ {g'}^2  (v^2_d - v^2_u) \delta_{\alpha\beta}}{4}  \, .
\label{eq:sleptonmassRR}
\end{eqnarray}
The indices $\alpha,\beta,\gamma=1,2,3$ characterize the flavours 
$e,\mu,\tau$, respectively.
$M^2_{L}$ and $M^2_{E}$ are the hermitean soft SUSY breaking mass matrices for
left and right sleptons, respectively. 
$A_{\alpha\beta}$ are the trilinear soft
SUSY breaking couplings of the sleptons and the Higgs boson:
${\mathcal L}_{\rm int}=-A_{\alpha\beta} \tilde\ell_{\beta R}^\dagger
\tilde\ell_{\alpha L} H^0_1+\cdots$.
$v_u$ and $v_d$ are the vacuum expectation values of the Higgs fields
with $v_u=\sqrt{2}~\langle H^0_2\rangle$, $v_d=\sqrt{2}~\langle H^0_1\rangle$,
and $\tan\beta\equiv v_u/v_d$. We work in a basis where the Yukawa coupling
matrix $Y^E_{\alpha\beta}$ of the charged leptons is real and flavour
diagonal with $Y^E_{\alpha\alpha}=\sqrt{2}~m_{\ell_\alpha}/v_d~
(\ell_\alpha=e,\mu,\tau)$.
The physical mass eigenstates $\tilde \ell_i$ are given by 
$\tilde \ell_i = R^{\tilde \ell}_{i\alpha} \tilde \ell'_\alpha$ with 
$\tilde\ell'_\alpha = (\tilde e_L, \tilde \mu_L, \tilde \tau_L,
\tilde e_R, \tilde \mu_R, \tilde \tau_R)$. 
The mixing matrix $R^{\tilde\ell}$ and the physical mass eigenvalues
are obtained by an unitary transformation 
$R^{\tilde\ell}M^2_{\tilde \ell}R^{\tilde\ell\dagger}=
{\rm diag}(m^2_{\tilde\ell_1},\dots,m^2_{\tilde\ell_6})$, 
where $m_{\tilde\ell_i} < m_{\tilde\ell_j}$ for $i<j$.
Similarly, one has for the sneutrinos in the basis of
$(\tilde\nu_{eL},\tilde\nu_{\mu L},\tilde\nu_{\tau L})\equiv
(\tilde\nu_{e},\tilde\nu_{\mu},\tilde\nu_{\tau})$
\begin{eqnarray}
M^2_{\tilde \nu,\alpha\beta} &=&  M^2_{L,\alpha\beta} 
+ \frac{\left( g^2 + {g'}^2 \right)(v^2_d - v^2_u)\delta_{\alpha\beta}}{8}
\qquad (\alpha,\beta=1,2,3)
\label{eq:sneutrinomass}
\end{eqnarray}
with the physical mass eigenstates 
$\tilde \nu_i = R^{\tilde \nu}_{i\alpha}\tilde \nu_\alpha'$ 
($m_{\tilde \nu_1} < m_{\tilde \nu_2} < m_{\tilde \nu_3}$) and 
$\tilde \nu_\alpha' = (\tilde \nu_e, \tilde \nu_\mu, \tilde \nu_\tau) $.
The relevant interaction Lagrangian for this study in terms of mass
eigenstates is given by:
\begin{eqnarray}
\label{eq:CoupChiSfermion}
{\mathcal L} &=& \bar \ell_i ( c^L_{ikm} P_L + c^R_{ikm} P_R)
           \tilde \chi^0_k \tilde \ell_m  
    + \bar{\ell_i} (d^L_{ilj} P_L + d^R_{ilj} P_R)
           \tilde \chi^-_l \tilde{\nu}_j
+\bar\nu_i e^R_{ikj}P_R \tilde \chi^0_k \tilde\nu_j
\nonumber \\
 &+& \bar{\nu_i} f^R_{ilm} P_R \tilde \chi^+_l \tilde{\ell}_m
+{\rm h.c.}~.
\end{eqnarray}
The specific forms of the couplings $c^L_{ikm}$, $c^R_{ikm}$,
$d^L_{ilj}$, $d^R_{ilj}$, $e^R_{ikj}$ and $f^R_{ilm}$ can be found in
\cite{Chung:2003fi}. The first two terms in \eq{eq:CoupChiSfermion}
give rise to the LFV signals studied here, whereas the 
last one will give rise to the SUSY background
because the neutrino flavour cannot be discriminated in high energy collider
experiments.

\section{Lepton flavour violating decays of sleptons and neutralinos}

Now we discuss systematically LFV decays of charged sleptons and neutralinos.
We will first consider cases where only two generations mix and afterwards
the case of three generation mixing. For definitness, we consider the study 
point SPS1a' \cite{spa}.
In this scenario we have a relatively light spectrum of charginos/neutralinos 
and sleptons with the three lighter charged sleptons being mainly $\tilde\ell_R$. 
This means that the flavour off-diagonal
elements of $M^2_{E,\alpha\beta}$ in Eq.~(\ref{eq:sleptonmassRR}) are 
expected to give the most important contribution to the LFV decays of 
the lighter charginos/neutralinos 
and sleptons. We therefore discuss LFV only in the right slepton sector.
For the SPS1a' point the relevant on-shell SUSY parameters are given by:
$\tan\beta=10$, $M_1=100.1$~GeV, $M_2=197.4$~GeV, $\mu=400$~GeV
$M_{L,11}=M_{L,22}=184$~GeV, $M_{L,33}=182.5$~GeV,  
$M_{E,11}=117.793$~GeV, $M_{E,22}=117.797$~GeV, $M_{E,33}=111$~GeV,
$A_{11}=-0.013$~GeV, $A_{22}=-2.8$~GeV, $A_{33}=-46$~GeV.
Here $M_1$ and $M_2$ are the $U(1)$ and $SU(2)$ gaugino mass
parameters, respectively
The mass spectrum of the lighter neutralinos and sleptons is: 
$m_{\tilde \chi^0_1}=97.8$~GeV, $m_{\tilde \chi^0_2}=184$~GeV, 
$m_{\tilde e_1}=125.251$~GeV, $m_{\tilde \mu_1}=125.212$~GeV, 
$m_{\tilde\tau_1}=107.4$~GeV. We have taken 
$M^2_{E,22}=M^2_{E,11}+1$~GeV$^2$, i.e. $M_{E,22}=M_{E,11}+4$~MeV, to 
avoid potential numerical problems 
in the definition of the LFV mixing angles (see Eq.~(\ref{eq:angle}) below). 
This results in a mass difference
$m_{\tilde e_1}-m_{\tilde \mu_1} = 39$~MeV which is well below
the expected mass resolution of LHC or ILC.
We consider the following $\tilde \chi^0_2$ decays
\begin{eqnarray}
\tilde \chi^0_2 ~\longrightarrow  ~ \ell^\pm_i \tilde \ell^\mp_j~ \,\, ,
\end{eqnarray}
where $\tilde\ell_1=\tilde\tau_1\sim\tilde{\tau}_R$, 
$\tilde \ell_2=\tilde \mu_1\simeq\tilde \mu_R$ and $\tilde \ell_3
=\tilde e_1\simeq\tilde e_R$.
They decay further as
\begin{eqnarray}
\tilde \ell^\mp_j~ \longrightarrow~ \ell^\mp_k \tilde \chi^0_1~,
\end{eqnarray}
see Table~\ref{tab:br}.\\
As the next step in our analysis, we add lepton flavour violating real
parameters $M^2_{E,\alpha\beta}$ with
$\alpha\ne \beta$ inducing LFV decays of neutralinos/charginos
and sleptons.
It is convenient to define the following effective lepton flavour
mixing angles
\begin{eqnarray}
\tan 2 \theta^{eff}_{\alpha\beta} \equiv 
\frac{2 M^2_{E,\alpha\beta}}{M^2_{E,\alpha\alpha} - M^2_{E,\beta\beta}}~,
\qquad (\alpha < \beta)
\label{eq:angle}
\end{eqnarray}
which are a measure of LFV.

\begin{figure}
\setlength{\unitlength}{1mm}
\begin{center}
\begin{picture}(140,80)
\put(-53,-114){\mbox{\epsfig{figure=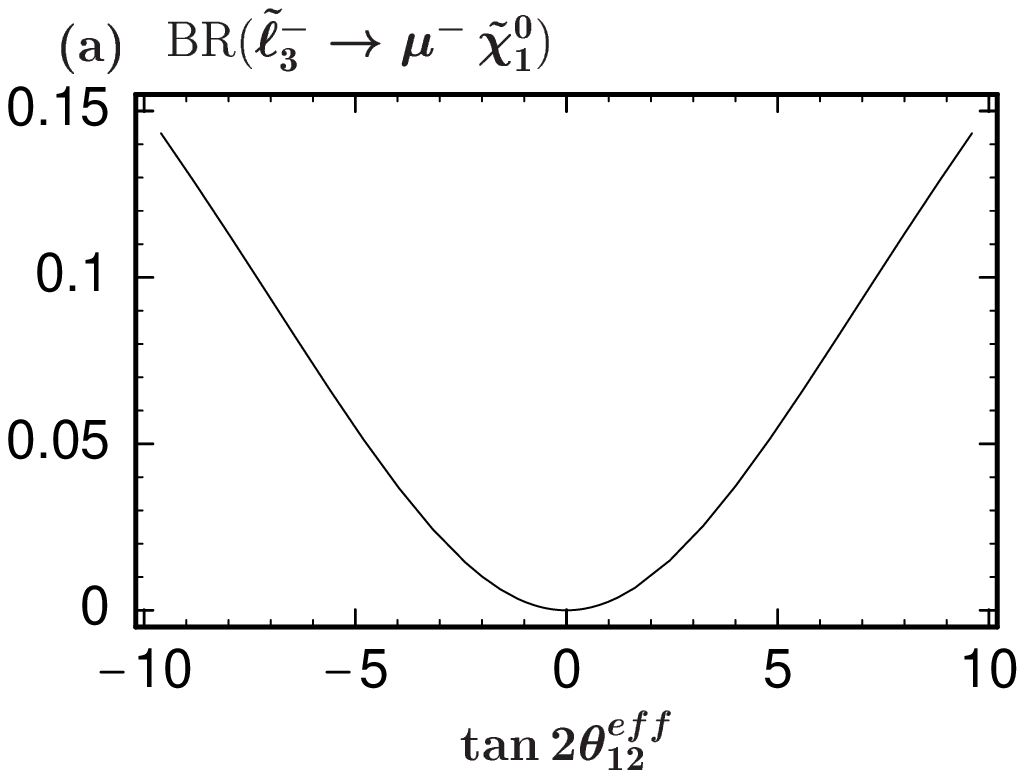,height=22.5cm,width=14.4cm}}}
\put(27,-115){\mbox{\epsfig{figure=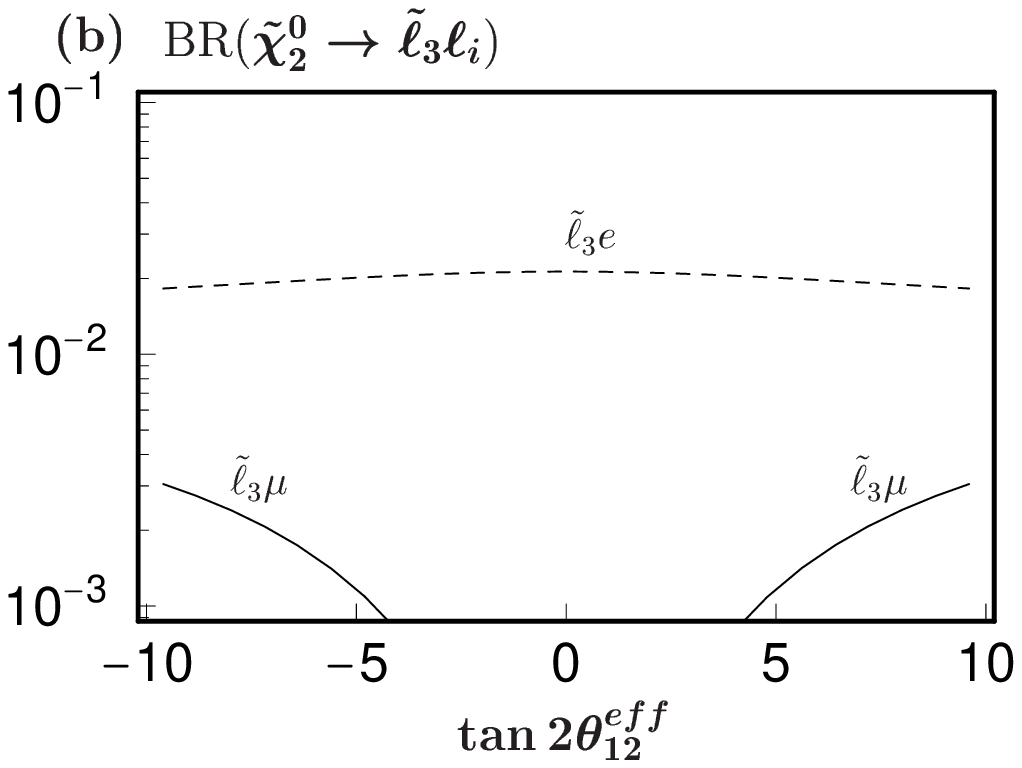,height=22.7cm,width=14.1cm}}}
\end{picture}
\end{center}
\vspace{-3cm}
\caption{In (a) we show  BR($\tilde\ell^-_3 \to \mu^- \tilde \chi^0_1 $)
as a function of $\tan 2 \theta^{eff}_{12}$ and in (b)
BR($\tilde \chi^0_2 \to \tilde \ell_3 \, \mu$) and 
BR($\tilde \chi^0_2 \to \tilde \ell_3 \, e$) as a function
of $\tan 2 \theta^{eff}_{12}$ summing over the charges, where
$\tilde\ell_3$ is dominantly $\tilde e_1$ with an admixture
of $\tilde\mu_R$.}
\label{fig:m12}
\end{figure}

In \fig{fig:m12} we show the lepton flavour violating branching ratios
BR($\tilde \ell^-_3 \to \mu^- \tilde \chi^0_1$) and 
BR($\tilde\chi^0_2\to\tilde\ell_3\ell_i$) as a function of 
$\tan 2 \theta^{eff}_{12}$. The parameter $M^2_{E,12}$ has been varied
in the full range satisfying BR$(\mueg) < 1.2\times 10^{-11}$, with
$M^2_{E,13}=M^2_{E,23}=0$.
Note that $\tilde\ell_3=\tilde e_1\simeq\tilde e_R$ for
$\tan 2 \theta^{eff}_{12}=0$ whereas $\tilde\ell_3$ is mainly 
$\tilde e_1$ with an admixture of $\tilde\mu_R$
for $\tan 2 \theta^{eff}_{12}\neq 0$.
As can be seen in \fig{fig:m12}(a) 
BR($\tilde \ell^-_3 \to \mu^- \tilde \chi^0_1$) can go up to $\sim$15\%.
As $\tilde\ell_2\sim \tilde\mu_1$ and $\tilde\ell_3\sim \tilde e_1$,
one has ${\rm BR}(\tilde \ell_3 \to \mu \tilde \chi^0_1)\simeq
{\rm BR}(\tilde \ell_2 \to e \, \tilde \chi^0_1)$ for 
a fixed value of $\tan 2 \theta^{eff}_{12}$. 
In \fig{fig:m12}(b) we show the LFV branching ratio
${\rm BR}(\tilde \chi^0_2 \to \tilde \ell_3 \mu)$ (full line); 
the LFV branching ratio ${\rm BR}(\tilde \chi^0_2 \to \tilde \ell_2 e)$ has 
the same value.
We also show the ``lepton flavour conserving (LFC)'' branching ratio
${\rm BR}(\tilde \chi^0_2 \to \tilde \ell_3 e)$ (dashed line);
the branching ratio ${\rm BR}(\tilde \chi^0_2 \to \tilde \ell_2 \mu)$ again has 
the same value.
The sum of the LFV branching ratios of $\tilde\chi^0_2$, 
${\rm BR}(\tilde \chi^0_2 \to \tilde \ell_3 \mu)$ and 
${\rm BR}(\tilde \chi^0_2 \to \tilde \ell_2 e)$, can reach about 0.6\%,
which is about $1/6$ of the sum of ``LFC''
branching ratios ${\rm BR}(\tilde \chi^0_2\to \tilde\ell_3 e)+
{\rm BR}(\tilde \chi^0_2\to \tilde\ell_2 \mu)$. 
Note that the dominant decay channels of $\tilde \chi^0_2$ are into
$\tilde\tau_1\tau$ and $\tilde\nu_{\ell}~\nu_{\ell}$
(see Table~\ref{tab:br}).

\begin{figure}[t]
\setlength{\unitlength}{1mm}
\begin{center}
\begin{picture}(140,90)
\put(-53,-110){\mbox{\epsfig{figure=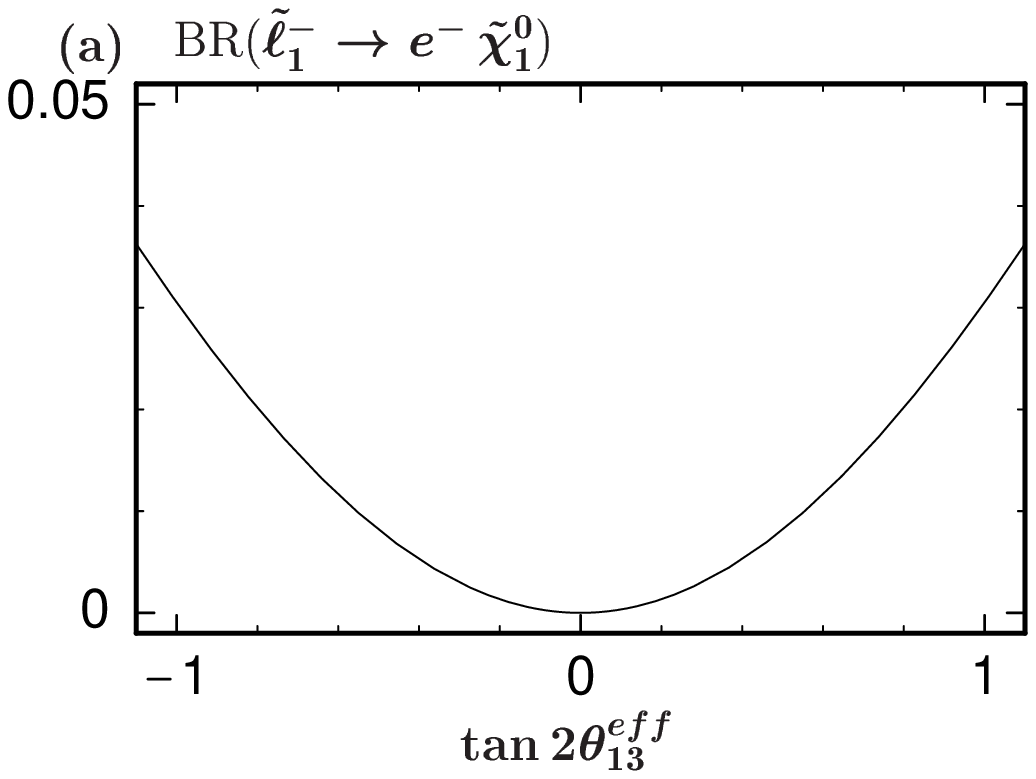,height=22cm,width=14.4cm}}}
\put(27,-110){\mbox{\epsfig{figure=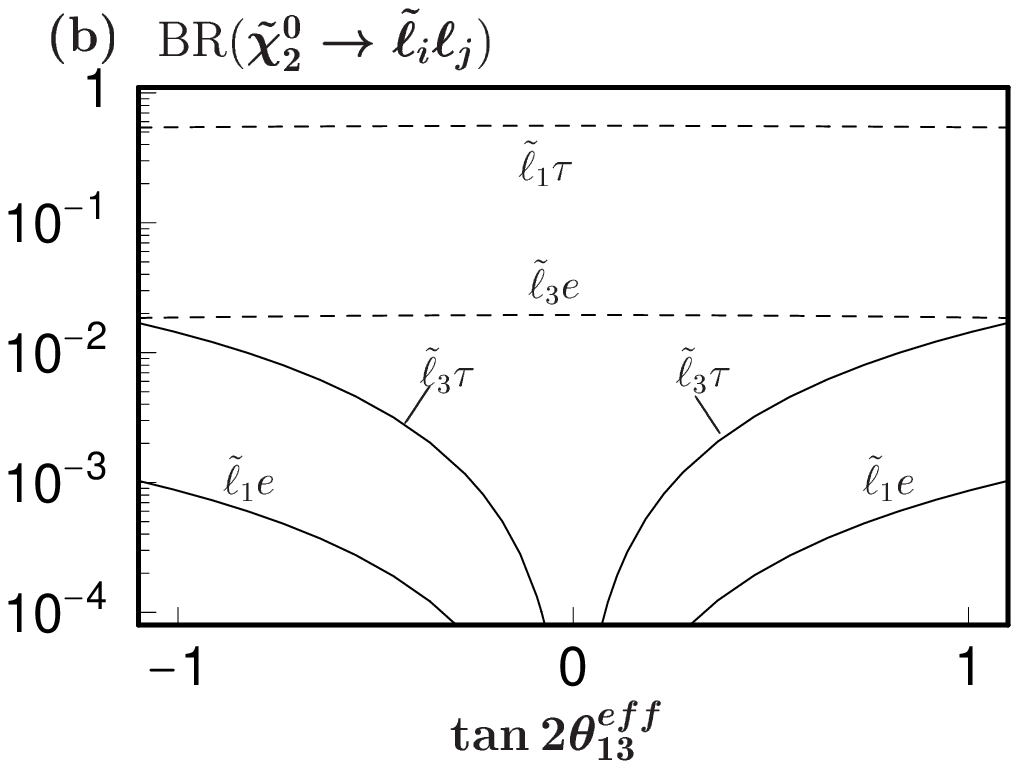,height=22.cm,width=14.1cm}}}
\end{picture}
\end{center}
\vspace{-3cm}
\caption{
In (a) we show  BR($\tilde\ell^-_1 \to e^- \tilde \chi^0_1 $)
as a function of $\tan 2 \theta^{eff}_{13}$ and in (b)
BR($\tilde \chi^0_2 \to \tilde \ell_3 \, \tau$),
BR($\tilde \chi^0_2 \to \tilde \ell_1 \, e$),
BR($\tilde \chi^0_2 \to \tilde \ell_1 \, \tau$)
and BR($\tilde \chi^0_2 \to \tilde \ell_3 \, e$) as 
a function of $\tan 2 \theta^{eff}_{13}$ summing over the charges, 
where $\tilde\ell_1$ ($\tilde\ell_3$) is dominantly 
$\tilde\tau_1$ ($\tilde e_1$) with an admixture
of $\tilde e_R$ ($\tilde\tau_R$).}
\label{fig:m13}
\end{figure}

In \fig{fig:m13} we take $M^2_{E,13} \ne 0$ varying it in the full range satisfying
BR$(\taueg) < 1.1 \times 10^{-7}$, with $M^2_{E,12}=M^2_{E,23}=0$.
In \fig{fig:m13}(a) we show the LFV branching ratio
BR($\tilde \ell^-_1 \to e^- \, \tilde \chi^0_1$) as a function of 
$\tan 2 \theta^{eff}_{13}$, where $\tilde\ell_1$ ($\tilde\ell_3$) is dominantly 
$\tilde\tau_1$ ($\tilde e_1$) with an admixture
of $\tilde e_R$ ($\tilde\tau_R$). 
The LFV branching ratio can go up to 3.5\%.
In \fig{fig:m13}(b) we plot the LFV branching ratios
BR($\tilde \chi^0_2 \to \tilde \ell_3 \, \tau$) and
BR($\tilde \chi^0_2 \to \tilde \ell_1 \, e$) as well
as the ``LFC'' branching ratios
BR($\tilde \chi^0_2 \to \tilde \ell_1 \, \tau$)
and BR($\tilde \chi^0_2 \to \tilde \ell_3 \, e$).
The LFV branching ratio BR($\tilde\chi_2\to\tilde\ell_3\tau$) 
can reach about 2\%.
The relative magnitudes of the branching ratios in \fig{fig:m13}(b)
are explained as follows: for the SPS1a' scenario we have 
$\tilde \chi^0_2\sim \tilde W^3$ so that
the $\tilde \chi^0_2$ couples strongly to $\tilde\ell_L$ and only
weakly to $\tilde\ell_R$. However,
only in the $\tilde\tau$-sector there is a significant $\tilde\ell_L-\tilde\ell_R$ 
mixing. Therefore, for large $\tan 2 \theta^{eff}_{13}$, the LFV branching ratio 
${\rm BR}(\tilde \chi^0_2 \to \tilde \ell_3 \tau)$ has about the
same size as the ``LFC'' branching ratio 
${\rm BR}(\tilde \chi^0_2 \to \tilde \ell_3 e)$.

We have also considered the case $M^2_{E,23}\neq 0$, $M^2_{E,12}=M^2_{E,13}=0$ for the
LFV decays $\tilde \ell^-_1 \to \mu^-\tilde \chi^0_1$, 
$\tilde\chi^0_2\to\tilde\ell_2\tau$ and 
$\tilde \chi^0_2 \to \tilde \ell_1 \mu$ 
($\tilde\ell_1\sim \tilde\tau_1$ and $\tilde\ell_2\sim \tilde\mu_1$).
In this case we have found similar behaviours as in the above case
($M^2_{E,13}\neq 0$, $M^2_{E,12}=M^2_{E,23}=0$) because the 
experimental bounds on BR($\taueg$) 
and BR($\taumug$) are similar.

\begin{figure}[t]
\setlength{\unitlength}{1mm}
\begin{center}
\begin{picture}(140,90)
\put(-53,-110){\mbox{\epsfig{figure=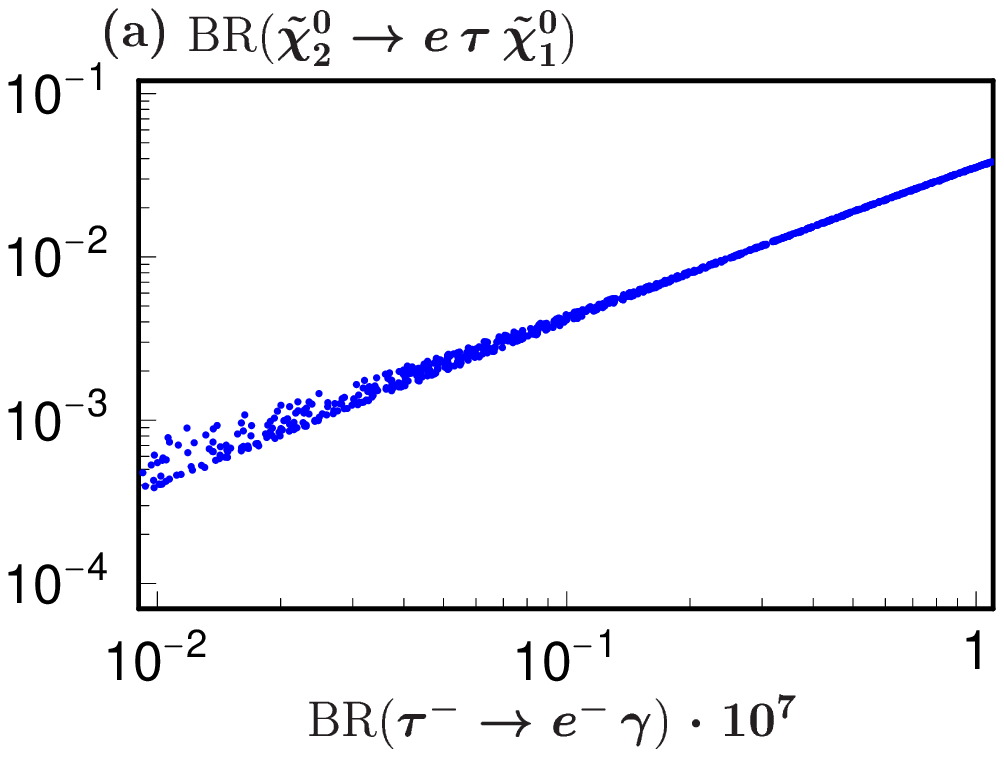,height=22.3cm,width=14.4cm}}}
\put(27,-108){\mbox{\epsfig{figure=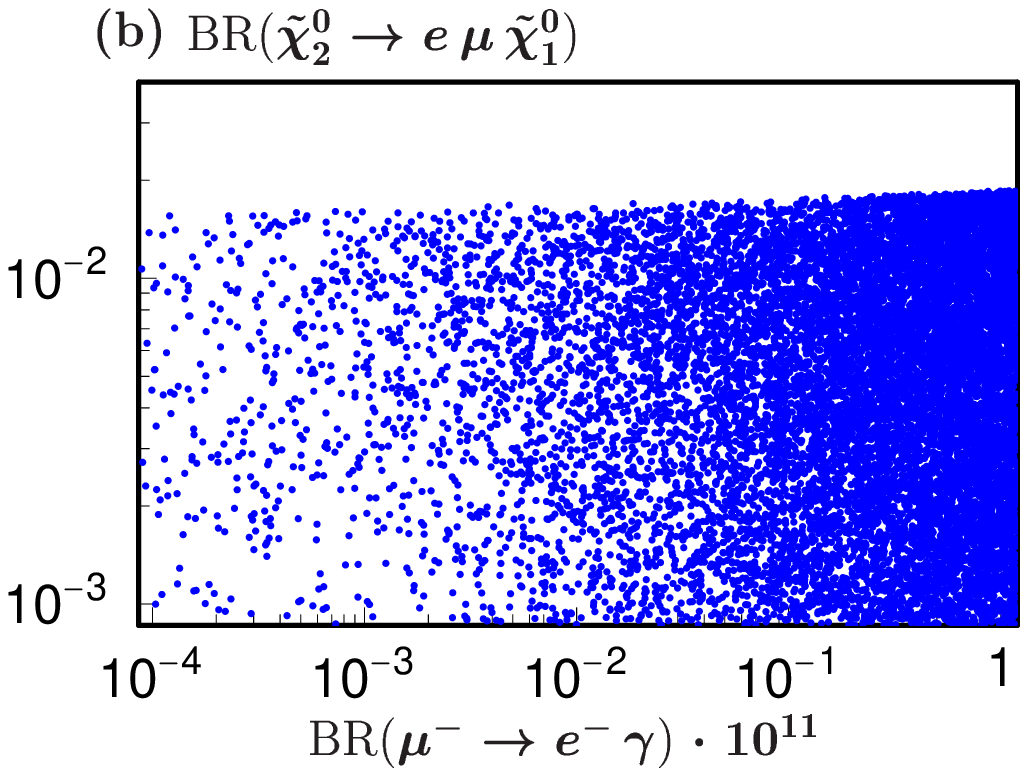,height=22cm,width=14.1cm}}}
\end{picture}
\end{center}
\vspace{-3.5cm}
\caption{LFV decay branching ratios BR($\tilde \chi^0_2\to e \tau \tilde{\chi}^0_1$)
and BR($\tilde \chi^0_2\to e \mu \tilde{\chi}^0_1$) 
as a function of BR($\taueg$) and BR($\mueg$), respectively, varying
the LFV parameters around the SPS1a' point.}
\label{fig:scatter}
\end{figure}

In \fig{fig:scatter}(a) we show 
BR($\tilde \chi^0_2 \to e~\tau~\tilde \chi^0_1$) summed over all intermediate
sleptons and all charges as a function of BR($\taueg$). 
The former branching ratio has been calculated by using the formula
\begin{eqnarray}
{\rm BR}(\tilde \chi^0_2 \to e~\tau~\tilde \chi^0_1) &=&
\sum_{i=1}^3\left[
{\rm BR}(\tilde \chi^0_2 \to e~\tilde\ell_i)
{\rm BR}(\tilde\ell_i \to \tau~\tilde \chi^0_1)\right.\nonumber\\[3mm]
&&{}
\left. +{\rm BR}(\tilde \chi^0_2 \to\tau~\tilde\ell_i)
{\rm BR}(\tilde\ell_i \to e~\tilde \chi^0_1)
\right]~.
\label{eq:branching}
\end{eqnarray}
We have randomly varied all off-diagonal entries in $M^2_E$ such that
all experimental constraints due to the rare lepton decays are fulfilled at the
same time. We see a strong correlation between
${\rm BR}(\tilde \chi^0_2 \to e~\tau~\tilde \chi^0_1)$ and
BR($\taueg$). This correlation appears since both
${\rm BR}(\tilde \chi^0_2 \to e~\tau~\tilde \chi^0_1)$ and
BR($\taueg$) depend strongly on the parameter $M_{E,13}$. 
We have found a similar strong correlation between
BR($\tilde \chi^0_2 \to \mu~\tau~\tilde \chi^0_1$) and BR($\taumug$).
In \fig{fig:scatter}(b) we show the branching ratio 
${\rm BR}(\tilde \chi^0_2 \to e~\mu~\tilde \chi^0_1)$ and find 
its upper bound of about $2\%$ almost
independent of BR($\mueg$). This independence
can be understood in the following way: In the mass insertion approximation,
there are two contributions to $\mueg$. The first one is due
to $\tilde{\mu}-\tilde{e}$ mixing and the second one is due to
the product of $\tilde{\mu}-\tilde{\tau}$ and 
$\tilde{\tau}-\tilde{e}$ mixings. 
As the constraints on rare $\tau$ decays are much less stringent 
than those on rare $\mu$ decays, the second one can be destructive and
as important as the first one.
This means that sizable ${\rm BR}(\tilde \chi^0_2 \to e~\mu~\tilde \chi^0_1)$
is possible even for smaller BR($\mueg$) in case of three generation mixing
differently from that of two generation mixing as studied in 
\cite{Agashe:1999bm,ref12,Hisano:2002tk} 
\footnote{Another possibility to enhance 
${\rm BR}(\tilde \chi^0_2 \to e~\mu~\tilde \chi^0_1)$ 
is to choose certain ratios between the higgsino mass parameter $\mu$ and
the gaugino mass parameter $M_2$ \cite{Hisano:2002tk}.}.
Note that,  in the SPS1a' scenario the masses of $\tilde\ell_2$ and 
$\tilde \ell_3$ ($\tilde\ell_2\simeq \tilde \mu_R$, 
$\tilde\ell_3\simeq \tilde e_R$)
are nearly degenerate and, hence, interference terms due to slepton flavour
oscillation may reduce ${\rm BR}(\tilde \chi^0_2 \to e~\mu~\tilde \chi^0_1)$
significantly \cite{ref8}. We have found numerically that the
formula in Eq.~(\ref{eq:branching}) with $\tau$ replaced by $\mu$
reproduces the correct results within 10\% error if 
$(m_{\tilde\ell_3}-m_{\tilde \ell_2})/\Gamma\geq 3$
[$\Gamma=\Gamma_{\tilde \ell_2}\simeq \Gamma_{\tilde \ell_3}$].

\section{Effects on di-lepton invariant mass spectra}

Now we consider LFV effects on the di-lepton mass
distribution in $\tilde\chi^0_2$ decays  
\begin{eqnarray}
\tilde \chi^0_2 \to \tilde \ell^+_i \ell^-_j \to \ell^+_k \ell^-_j \tilde \chi^0_1~.
\label{eq:chi2dec}
\end{eqnarray}
These decays can appear in the cascade decays of squarks and gluinos
as produced at LHC. In these events one studies the invariant
di-lepton mass spectrum 
$d N / d m(\ell \ell)$ with $m(\ell\ell)^2 = (p_{\ell^+} + p_{\ell^-})^2$. 
Its kinematical endpoint is used in
combination with other observables to determine masses or
mass differences of sparticles \cite{Paige:1996nx,Bachacou:1999zb,Allanach:2000kt}. 
These spectra will change in the presence of lepton flavour violation.

\begin{figure}[t]
\begin{center}
\setlength{\unitlength}{1mm}
\begin{picture}(150,80)
\put(0,0){\epsfig{figure=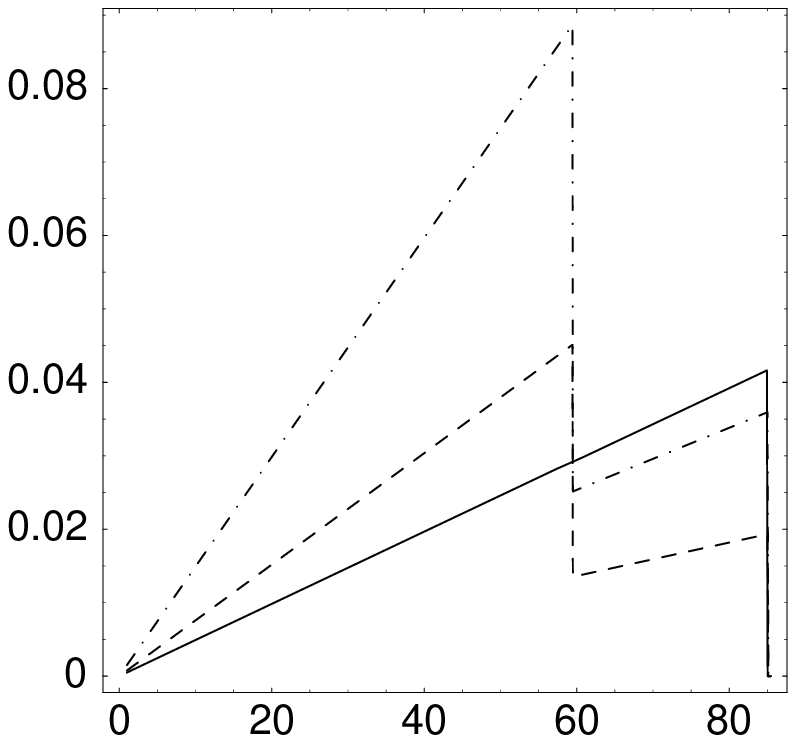,height=70mm,width=70mm}}
\put(-12,71){\mbox{\bf (a)} 
{$100 \Gamma^{-1}_{tot}
  d \, \Gamma(\tilde \chi^0_2 \to \ell^\pm_i \ell^\mp_j \tilde \chi^0_1) 
               / d \, m(\ell^\pm_i \ell^\mp_j)$~[GeV$^{-1}$]}}

\put(45,-1){\mbox{$m(\ell^\pm_i \ell^\mp_j)$~[GeV]}}
\put(32,46){\mbox{$e\tau$}}
\put(30,26){\mbox{$\mu\tau$}}
\put(60,34){\mbox{$e\mu$}}
\put(82,0){\epsfig{figure=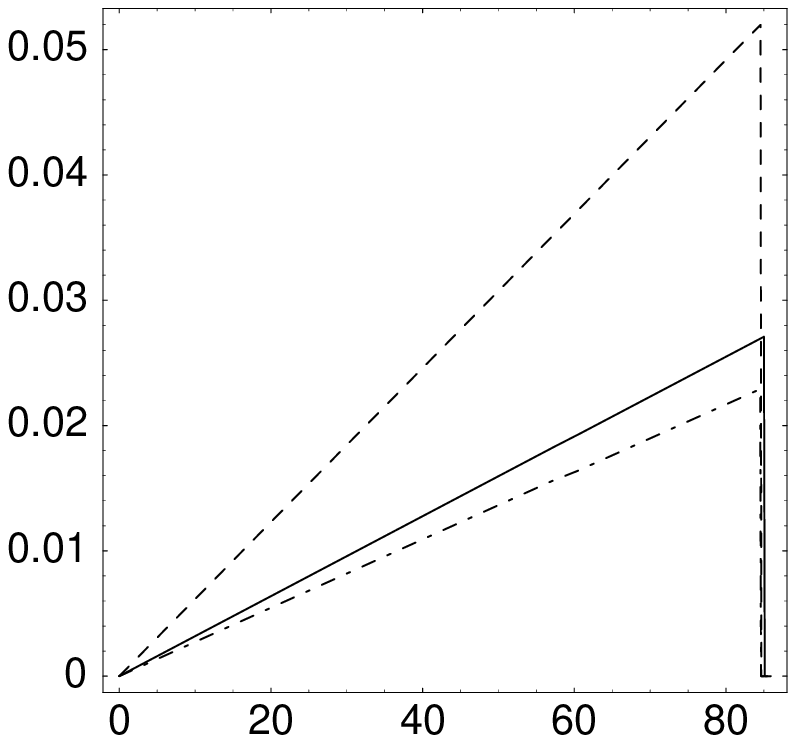,height=70mm,width=70mm}}
\put(80,71){\mbox{{\bf (b)} $100 \Gamma^{-1}_{tot}
  d \, \Gamma(\tilde \chi^0_2 \to \ell^+ \ell^- \tilde \chi^0_1) 
               / d \, m(\ell^+ \ell^-)$~[GeV$^{-1}$]}}
\put(126,-1){\mbox{$m(\ell^+ \ell^-)$~[GeV]}}
\put(127,52){\mbox{$\mu^+\mu^-$}}
\put(137,61){\mbox{$e^+ e^-$}}
\put(130,33){\mbox{$\mu^+\mu^-$}}
\put(131,18){\mbox{$e^+ e^-$}}
\put(131,20){\line(-2,3){2.5}}
\end{picture}
\end{center}
\caption{Invariant mass spectra 
$100\Gamma^{-1}_{tot} d \Gamma(\tilde \chi^0_2 \to \ell_i \ell_j \tilde \chi^0_1) 
               / d m(\ell_i \ell_j)$ versus $m(\ell_i \ell_j)$.
In {\bf (a)} we show
the ``flavour violating'' spectra summed over charges in the LFV case for
the SPS1a' scenario:
 $e^\pm \mu^\mp$ (full line), $e^\pm \tau^\mp$ (dashed dotted line)
and $\mu^\pm \tau^\mp$ (dashed line) and in 
{\bf (b)}  we show the ``flavour conserving'' spectra: 
$e^+ e^-$ (dashed line) and $\mu^+ \mu^-$ (dashed line) are
for the LFC case in the SPS1a' scenario, and 
$e^+ e^-$ (dashed dotted line) and 
$\mu^+ \mu^-$ (full line) are for the LFV case in the SPS1a' scenario.}
\label{fig:dilepton}
\end{figure}

To illustrate the effect of LFV on these spectra, in \fig{fig:dilepton} we present 
invariant mass distributions for various lepton pairs taking
the following LFV parameters: $M^2_{E,12} = 30$~GeV$^2$, 
$M^2_{E,13} = 850$~GeV$^2$ and $M^2_{E,23} = 600$~GeV$^2$,
for which we have $(m_{\tilde\ell_1},m_{\tilde\ell_2},m_{\tilde\ell_3})$ $=
(106.4,125.1,126.2)$~GeV.
These parameters are chosen such that large LFV $\tilde\chi^0_2$ 
decay branching ratios are possible consistently with the experimental 
bounds on the rare lepton decays (see \fig{fig:scatter}). 
For this set of parameters we obtain BR$(\mueg)=9.5 \times 10^{-12}$,
BR$(\taueg)=1.0 \times 10^{-7}$ and BR$(\taumug)=5.2 \times 10^{-8}$.
In Table~\ref{tab:br} we show the slepton and $\tilde\chi^0_2$ decay
branching ratios in the LFV case (as well as in the lepton flavour conserving (LFC) case).
By using Table~\ref{tab:br} and the formulae analogous to Eq.~(\ref{eq:branching})
we obtain the following $\tilde\chi^0_2$ decay branching ratios in this
LFV case: BR$(e\mu)=1.7$\%, BR$(e\tau)=3.4$\%, BR$(\mu\tau)=1.8$\%, BR$(e^+ e^-)=1$\%, 
BR$(\mu^+\mu^-)=1.2$\%, BR$(\tau^+\tau^-)=51$\% with BR$(\ell_i\ell_j)\equiv 
{\rm BR}(\tilde\chi^0_2\to \ell_i\ell_j\tilde\chi^0_1)$.
In \fig{fig:dilepton}(a) we show the flavour violating spectra
$(100/\Gamma_{tot}) d \Gamma(\tilde \chi^0_2 \to \ell^\pm_i \ell^\mp_j \tilde \chi^0_1) 
/ d m(\ell^\pm_i \ell^\mp_j)$ versus $m(\ell^\pm_i \ell^\mp_j)$ for the final states
$\mu \tau$, $e \tau$ and $e \mu$.
In cases where the final state contains a $\tau$-lepton, 
one finds two sharp edges. The
first one at $m \simeq 59.4$~GeV is due to an intermediate 
$\tilde \ell_1 (\sim \tilde \tau_R)$ 
and the second one at $m \simeq 84.6$~GeV is due to intermediate states of
the two heavier sleptons $\tilde\ell_2~(\sim \tilde\mu_R)$ and  
$\tilde\ell_3~(\sim \tilde e_R)$ with $m_{\tilde\ell_2}\simeq m_{\tilde\ell_3}$
(see Eq.~(\ref{eq:branching})). The position of the 
edges can be expressed in terms of the neutralino and intermediate slepton 
masses \cite{Paige:1996nx}:
\begin{eqnarray}
m^2_{edge}(\ell\ell) = \frac{(m^2_{\tilde \chi^0_2} - m^2_{\tilde\ell_i})
                       ( m^2_{\tilde\ell_i} - m^2_{\tilde \chi^0_1})}
                      {m^2_{\tilde\ell_i}}~.
\label{eq:egde}
\end{eqnarray}
In the case of the $e \mu$ spectrum the first edge is practically invisible
because the branching ratios of $\tilde \chi^0_2$ into $\tilde\ell_1 \, e$
and $\tilde\ell_1 \, \mu$ are tiny as can be seen in Table~\ref{tab:br}. 
Note that the rate for the $e \tau$ final state is largest in our case because 
$|M^2_{E,13}|$ is larger than the other LFV parameters.

In \fig{fig:dilepton}(b) we show the ``flavour conserving'' spectra for the
final states with $e^+ e^-$ and $\mu^+ \mu^-$. The dashed line corresponds to 
the flavour conserving case where $M^2_{E,ij}=0$ for $i\ne j$.
LFV causes firstly a reduction of the height of the end point peak.
Secondly, it induces a difference between 
the $\mu^+ \mu^-$ and $e^+ e^-$ spectra
because the mixings among the three slepton generations are in 
general different from one another.
The peaks at $m \simeq 59.4$~GeV in the $\mu^+ \mu^-$ and $e^+ e^-$ spectra
are invisible as in the $e\mu$ spectrum, for the same reason as mentioned above.
As for the $\tau^+ \tau^-$ spetrum we remark that the height of the peak
(due to the intermediate $\tilde\ell_1$ ($\sim \tilde\tau_R$))
in the $\tau^+ \tau^-$ spetrum gets reduced by about 5\% and that the
contributions due to the intermediate $\tilde\ell_{2,3}$ are invisible.
Moreover, the peak position gets shifted to a smaller value by about 2.7 GeV since
the mass of the intermediate $\tilde\ell_1$ gets reduced by 1 GeV 
compared to the flavour conserving case. 

It is interesting to note that in the LFV case the rate of the channel 
$e\tau$ can be larger than those of the channels with the 
same flavour, $e^+ e^-$ and $\mu^+ \mu^-$. Moreover,
by measuring all di-lepton spectra for the flavour violating as well as
conserving channels, one can make an important cross check
of this LFV scenario:
the first peak position of the lepton flavour violating spectra 
(except the $e\mu$ spectrum) must coincide
with the end point of the $\tau^+ \tau^-$ spectrum and the second
peak must coincide with those of the $e^+ e^-$ and $\mu^+ \mu^-$
spectra.

Up to now we have investigated in detail the di-lepton mass spectra
taking SPS1a' as a specific example. In the following we discuss
which requirements other scenarios must fulfill to observe double-edge
structures. Obviously the kinematic condition
$m_{\tilde\chi^0_s}> m_{\tilde\ell_i,\tilde\ell_j}>m_{\tilde\chi^0_r}$
must be fulfilled and sufficiently many  $\tilde\chi^0_s$ must be produced.
In addition there should be two sleptons contributing in a sizable 
way to the decay $\tilde\chi^0_s\to \ell' \ell''\tilde\chi^0_r$
and, of course, the corresponding branching ratio has to be large enough
to be observed. For this the corresponding LFV entries in the slepton
mass matrix have to be large enough. Moreover, also the mass difference between
the two contributing sleptons has to be sufficiently large so that the
difference of the positions of the two peaks is larger than the experimental 
resolution. In mSugra-like scenarios the kinematic requirements 
(including the positions of the peaks) are fulfilled in the regions
of parameter space where $m_0^2 \lsim 0.4~m^2_{1/2}$ and $\tan\beta\gsim 8$.
The first condition provides for right sleptons lighter than the $\tilde\chi^0_2$
and the second condition ensures that the mass difference between
$\tilde\tau_1$ and the other two right sleptons is sufficiently large.
In the region where $m_0^2 \lsim 0.05~m^2_{1/2}$ also the left sleptons
are lighter than $\tilde\chi^0_2$, giving the possibility of additional 
structures in the di-lepton mass spectra. We remark that SPS1a' is not 
the most favourable case because of the appearance of the decay 
$\tilde\chi^0_2\to \nu \nu \tilde\chi^0_1$ 
which has quite a large branching ratio ($\sim 40\%$). For example
in the original SPS1a point this decay mode is absent allowing for much larger
LFV branching ratios of $\tilde\chi^0_2$.

Finally we briefly discuss background reactions in the LFV search at LHC.
The largest SM background is due to $t \bar{t}$ production. 
There is also SUSY background due to uncorrelated 
leptons stemming from different squark and gluino decay
chains. The resulting di-lepton mass distributions will, however, be smooth and
decrease monotonically with increasing di-lepton invariant mass as was explicitly 
shown in a Monte Carlo analysis in \cite{ref12,Hisano:2002tk}.
It was also shown that the single edge structure can be observed over
the smooth background in the $e\mu$ and $\mu\tau$ invariant mass distributions.
Therefore the novel distributions as shown in \fig{fig:dilepton}, in particular
the characteristic double-edge structures in the $e\tau$ and $\mu\tau$ invariant mass
distributions, should be clearly visible on top of the background.
Note that the usual method for background suppression, by taking the sum 
$N(e^+ e^-)+N(\mu^+ \mu^-)-N(e^\pm \mu^\mp)$, is not
applicable in the case of LFV searches. Instead one has to study the individual 
pair mass spectra. Nevertheless, one can expect that these peaks will 
be well observable \cite{Hinchliffepc}.
Here note also that the tau lepton could be identified by a hadronic
``tau jet'' though the double-edge structures in the $e\tau$ and $\mu\tau$
spectra would get a smearing effect due to a neutrino emission \cite{ref12}. 
To show more clearly the observability of such LFV signals a detailed Monte Carlo
study would be necessary. This, however, is beyond the scope 
of the present paper. 

\begin{table}[t]
\begin{center}
\begin{tabular}{|c|c|c||c|c|c|}
\hline
channel & LFC case & LFV case &
channel & LFC case & LFV case \\ \hline
$\tilde \ell_1 \to \tilde \chi^0_1 \, e$ & 0 & 0.034 &
$\tilde \chi^0_2 \to \tilde \ell_1^\pm e^\mp$ & 0 & 0.001 \\
$\tilde \ell_1 \to \tilde \chi^0_1 \, \mu$ & 0 & 0.017 &
$\tilde \chi^0_2 \to \tilde \ell_1^\pm \mu^\mp$ & 0 & 0.0005 \\
$\tilde \ell_1 \to \tilde \chi^0_1 \, \tau$ & 1 & 0.949 &
$\tilde \chi^0_2 \to \tilde \ell_1^\pm \tau^\mp$ & 0.558 & 0.535 \\  \cline{1-3}
$\tilde \ell_2 \to \tilde \chi^0_1 \, e$ & 0 & 0.35 &
$\tilde \chi^0_2 \to \tilde \ell_2^\pm e^\mp$ & 0 & 0.007 \\
$\tilde \ell_2 \to \tilde \chi^0_1 \, \mu$ & 1 & 0.649 &
$\tilde \chi^0_2 \to \tilde \ell_2^\pm \mu^\mp$ & 0.021 & 0.014 \\
$\tilde \ell_2 \to \tilde \chi^0_1 \, \tau$ & 0 & 0.00002 &
$\tilde \chi^0_2 \to \tilde \ell_2^\pm \tau^\mp$ & 0 & 0.00001 \\ \cline{1-3}
$\tilde \ell_3 \to \tilde \chi^0_1 \, e$ & 1 & 0.62 &
$\tilde \chi^0_2 \to \tilde \ell_3^\pm e^\mp$ & 0.019 & 0.0117 \\
$\tilde \ell_3 \to \tilde \chi^0_1 \, \mu$ & 0 & 0.335 &
$\tilde \chi^0_2 \to \tilde \ell_3^\pm \mu^\mp$ & 0 & 0.0069 \\
$\tilde \ell_3 \to \tilde \chi^0_1 \, \tau$ & 0 & 0.044 &
$\tilde \chi^0_2 \to \tilde \ell_3^\pm \tau^\mp$ & 0 & 0.0234 \\ \cline{1-3}
\multicolumn{3}{c|}{} & $\tilde \chi^0_2 \to \tilde \nu \nu$ & 0.401 & 0.401 
\\ \cline{4-6}
\end{tabular}
\end{center}
\caption{Branching ratios of $\tilde \ell_{1,2,3}$ and
$\tilde\chi_2^0$ decays in the LFC and LFV cases for the SPS1a' scenario.
The LFV case is characterized by the following LFV parameters: 
$M^2_{E,12} = 30$~GeV$^2$, $M^2_{E,13} = 850$~GeV$^2$ 
and $M^2_{E,23} = 600$~GeV$^2$, for which one has 
$(m_{\tilde\ell_1},m_{\tilde\ell_2},m_{\tilde\ell_3})=(106.4,125.1,126.2)$~GeV. 
In the LFC case (i.e. in the case where $M^2_{E,12}=M^2_{E,13}=M^2_{E,23}=0$)
one has $(m_{\tilde\ell_1},m_{\tilde\ell_2},m_{\tilde\ell_3})
=(107.4,125.2,125.3)$~GeV. Note that 
$(\tilde\ell_1,\tilde\ell_2,\tilde\ell_3)=(\tilde\tau_1,\tilde\mu_1,\tilde e_1)$
in the LFC case and that 
$(\tilde\ell_1,\tilde\ell_2,\tilde\ell_3)\sim(\tilde\tau_R,\tilde\mu_R,\tilde e_R)$
in the LFV case.
\label{tab:br}}
\end{table}

\section{Summary}

To summarize, we have studied the effect of SUSY lepton flavour 
violation on the decay chains 
$\tilde\chi^0_2\to\ell^\mp_i\tilde\ell^\pm_j
\to \ell^\mp_i\ell^\pm_k\tilde\chi^0_1$, 
which may arise from cascade decays of gluinos and squarks at LHC.
As an example, we have adopted the SPS1a' scenario supplemented with
lepton flavour violating entries in the soft SUSY breaking mass matrix $M^2_E$, with
two and three generation mixings in the right slepton sector which
give the most important contributions to the LFV decays.
Additional mixings in the left-left and/or left-right sectors do not 
lead to a significant change of the LFV signals.
 
We have found that the most recent experimental bounds on flavour violating 
lepton decays allow for sizable flavour violating $\tilde\chi^0_2$ decay branching ratios
with the following upper limits:
BR($\tilde \chi^0_2 \to e~\mu~\tilde \chi^0_1$)$\lsim$ 2\%,
BR($\tilde \chi^0_2 \to e~\tau~\tilde \chi^0_1$)$\lsim$ 4\% and
BR($\tilde \chi^0_2 \to \mu~\tau~\tilde \chi^0_1$)$\lsim$ 3\%.
Moreover, a strong correlation between the branching ratios
BR($\tilde \chi^0_2 \to e~\tau~\tilde \chi^0_1$) 
(BR($\tilde \chi^0_2 \to \mu~\tau~\tilde \chi^0_1$)) and
BR($\taueg$) (BR($\taumug$)) is found. 
This would imply that if BR($\taueg$) or BR($\taumug$) were
measured not to be much below the current upper bound, then the  
signals of the corresponding lepton flavour violating 
neutralino decays should also be accessible at future collider experiments.
Furthermore, a sizable BR($\tilde \chi^0_2 \to e~\mu~\tilde \chi^0_1$)
can be compatible with a small BR($\mueg$) in case of three
generation mixing differently from two generation mixing cases
as previously studied.

In particular, we have studied the impact of LFV
due to three slepton generation mixing on the di-lepton mass
distributions from the decays 
$\tilde\chi^0_2\to \tilde\ell \ell'\to \ell' \ell''\tilde\chi^0_1$
measured at LHC. For the di-lepton spectra of two leptons with
equal flavour we have found that LFV leads to a reduction of the height of the
end-point peaks. This reduction is different for $e^+e^-$, $\mu^+\mu^-$ 
and $\tau^+\tau^-$ channels. This means, for example, that even in case
of nearly degenerate masses of $\tilde e_R$ and $\tilde\mu_R$ the
$e^+e^-$ mass spectrum can be significantly different from that of $\mu^+\mu^-$
due to three slepton generation mixing. 
For two leptons of different flavours we have found that novel and
characteristic edge structures in the distributions, such as a double-edge
structure in the $e\tau$ and $\mu\tau$ mass spectra can appear.
The double-egde structure stems from the mass difference between
$\tilde\tau_1 (\sim\tilde\tau_R)$ and $\tilde e_R, \tilde\mu_R$.
The appearance of such remarkable structures provides a powerful test of 
SUSY lepton flavour violation at LHC and useful informations on the flavour
structure of the slepton sector can be obtained. 
In such a case the additional peak may allow for a more precise measurement
of the mass of $\tilde\ell_1 (\sim\tilde\tau_1)$.
Finally, we have also worked out the conditions for the appearance of such
a double-edge structure in the di-lepton mass distributions for scenarios different 
from the SPS1a' scenario.


\section*{Acknowledgments}
%
We thank I.~Hinchliffe for useful discussions.
This work is supported by the `Fonds zur
F\"orderung der wissenschaftlichen Forschung' (FWF) of Austria, project
No. P16592-N02 and by Acciones Integradas Hispano--Austriaca.
W.P.~is supported by a MCyT Ramon y Cajal contract,
by the Spanish grant BFM2002-00345, by the
European Commission Human Potential Program RTN network
HPRN-CT-2000-00148  and partly by the Swiss 'Nationalfonds'.


\end{document}